\documentclass[aps,pra,reprint,amsmath,twoside,showpacs]{revtex4-1}
\usepackage{xspace}
\usepackage{graphicx}
\usepackage{multirow}
\usepackage{dcolumn}
\usepackage{braket}

\newcommand{\rmat}{\textit{R}-matrix\xspace}

\newcommand{\bspls}{\textit{B}-splines\xspace}
\newcommand{\seone}{$^1S^e$\xspace}
\newcommand{\sethree}{$^3S^e$\xspace}
\newcommand{\poone}{$^1P^o$\xspace}
\newcommand{\pothree}{$^3P^o$\xspace}
\newcommand{\deone}{$^1D^e$\xspace}
\newcommand{\dethree}{$^3D^e$\xspace}
\newcommand{\twos}{2$s$\xspace}
\newcommand{\twop}{2$p$\xspace}
\newcommand{\threes}{3$s$\xspace}
\newcommand{\abini}{\textit{ab initio}\xspace}
\newcommand{\figab}{Fig.~}
\newcommand{\eqab}{Eq.~}
\newcommand{\eqsab}{Eqs.~}
\newcommand{\refab}{Ref.~}
\newcommand{\refsab}{Refs.~}
\newcommand{\tabab}{Table~}
\newcommand{\secab}{Section~}
\newcommand{\electron}{e$^-$\xspace}
\begin{document}
	\title{\rmat calculations of electron collisions with lithium atom at low energies}
	\author{\firstname{Michal} Tarana}
	\author{\firstname{Roman} \v{C}ur\'ik}
	\affiliation{J. Heyrovsk\'y Institute of Physical Chemistry of the ASCR,~v.v.i., Dolej\v{s}kova 2155/3, 182 23 Prague 8, Czech Republic}
	\email{michal.tarana@jh-inst.cas.cz}
	\begin{abstract}
		\rmat calculations of the electron collisions with lithium atom at energies below 
		the \threes excitation threshold are presented. The \seone, \sethree 
		and \poone phase shifts calculated in the near-threshold energy range are in excellent 
		agreement with previous theoretical studies. The threshold behavior of the \pothree phase 
		shift is accurately analyzed along with the resonance located at the scattering energy 
		$\sim60$~meV. The phase shifts and cross sections calculated here show two resonances below 
		the \threes threshold that have not been previously reported.
	\end{abstract}
	\pacs{34.80.Bm, 34.80.Dp}
	\maketitle
	\section{Introduction}
	Ultra-long-range Rydberg molecules, for the first time theoretically predicted by 
	\citet{Greene-prl}, are very exotic systems in which one atom in its ground state interacts 
	with another atom in its highly excited Rydberg state with the distance of the nuclei varying 
	between $10^2$ and $10^4$ a.u \cite{Greene-prl,Khuskivadze,Hamilton2002}. The existence and 
	character of the electronic bound states of these molecules is determined by the low-energy 
	interaction between the Rydberg electron and the neutral atom in the ground state. Typically, 
	this interaction is approximated by the $s$-wave zero-range Fermi pseudopotential 
	\cite{fermi1934} and its $p$-wave extension \cite{Omont1977,Hamilton2002} or by the 
	finite-range model potential \cite{Khuskivadze}. Both models are constructed using the $s$-wave 
	and $p$-wave phase shifts of the corresponding electron-atom scattering process at energies below the 
	lowest threshold of the electronic excitation.
	
	So far, the ultra-cold quantum gases, particularly those consisting of the heavier alkali 
	metals, have provided the most suitable environment for the experimental realization and study 
	of the ultra-long-range Rydberg molecules \cite{Bendkowsky2009,Bellos,Bendkowsky-prl}. The design and 
	interpretation of these experiments requires accurate theoretical models of the long-range 
	Rydberg molecules and, therefore, accurate phase shifts of the electron collisions with the 
	alkali-metal atoms at the low scattering energies \cite{Bahrim2001}. Recently, \citet{Schmid2018} 
	proposed an experiment to study the ion-atom scattering in the ultracold regime based on the 
	photoionization of the Li-Li ultra-long-range Rydberg molecules. Although the low-energy 
	\electron-Li scattering has been studied both theoretically 
	\cite{Burke1969,Norcross1971,Moores1986,BuckmanClark} and experimentally 
	\cite{Leep1974,Jaduszliwer1981,BuckmanClark}, 
	the demand for the accurate and consistent data by the experimental research groups dealing 
	with the ultra-long-range Rydberg molecules involving lithium justifies 
	us to revisit this topic using very accurate contemporary computational methods.
	
	The \abini calculations by \citet{Norcross1971} provide very accurate 
	characterization of the \seone 
	and \sethree electron 
	collisions with the lithium atoms at very low scattering 
	energies between 0.1~meV and 68~meV. The  phase shifts calculated in this energy range are 
	fitted to the modified effective range theory (MERT) \cite{OMalley1961} and the accurate values 
	of the singlet and triplet scattering lengths are obtained.
	
	However, \citet{Norcross1971} calculated the \poone and \pothree scattering phase shifts only 
	for three values 
	of the scattering energies between 0.13~eV and 0.4~eV. Although the extrapolation of the \poone 
	phase shifts towards very low energies using the MERT is adequate, it is questionable in the 
	\pothree case since the lowest \pothree resonance is located below the interval where the 
	scattering calculations were performed.
	
	The low-energy \electron-Li scattering was also studied by \citet{Burke1969} using the 
	close-coupling (CC) expansion where the states of the neutral target were approximated by the 
	Hartree-Fock wave functions. The ranges of the scattering energies at which the phase shifts 
	are calculated in 
	\cite{Norcross1971} and \cite{Burke1969} overlap between 0.1~eV and 0.9~eV. Although the 
	\seone, \poone and \pothree phase shifts calculated by \citet{Burke1969} are in excellent 
	agreement with those published by \citet{Norcross1971} in this energy interval, their \sethree 
	phase shift raises more rapidly with decreasing scattering energy than in \refab\cite{Norcross1971}. 
	As a result, each of these two works \cite{Norcross1971,Burke1969} predicts the Ramsauer-Townsend minimum 
	at different energies. Moreover, the low-energy 
	\pothree phase shift and cross section published in \citet{Burke1969} show a clear resonance at 
	$\sim0.06$~eV. However, the character of the cross section below this resonance suggests that 
	the calculations by \citet{Burke1969} yield different threshold behavior than that predicted by 
	\citet{Norcross1971}.
	
	The experimental research of the electron-atom scattering becomes increasingly more challenging 
	with 
	decreasing collision energies. \citet{Jaduszliwer1981} measured the total \electron-Li 
	scattering cross 
	section above the \twop threshold. The excitation cross sections were measured by 
	\citet{Leep1974}. However, to our best knowledge, no experimental results have been published 
	for the scattering 
	energies below the lowest excitation threshold. Therefore, in order to 
	compare the present calculations with the experiment, it was necessary to 
	perform the \rmat computations for the energies above the \twop excitation threshold. Another 
	theoretical study in this energy region was published by \citet{Moores1986} who utilized the CC approach involving five 
	lowest states of the target.
	
	The goal of this paper is to introduce such a model of the \electron-Li collisions that provides 
	accurate 
	results from very low scattering energies to the \threes threshold of the electronic 
	excitation. 
	Parametrization of the phase shifts at very low scattering energies presented in this paper 
	provides the data 
	necessary for the research of the ultra-long range Rydberg molecules and other phenomena where 
	the electrons interact with the neutral lithium atom at low energies. Extension of the 
	calculations towards the energies above the \twop excitation threshold uncovers new resonances 
	that were not mentioned in the previously published papers. The reason 
	why they do not appear in the previously published studies \cite{Moores1986,Leep1974} is that 
	the 
	energy grids at which the cross sections were calculated \cite{Moores1986} and measured 
	\cite{Leep1974,Jaduszliwer1981} were not fine enough to resolve the corresponding narrow structures.
	
	The atomic units are used throughout the paper unless stated otherwise. Since 
	lithium is a very light element, no spin-orbit interaction or other relativistic effects are 
	considered in this work. The rest of this paper is organized as follows: 
	\secab\ref{sec:modelpot} deals with the representation of the Li$^+$ core by a model potential, 
	the parameters of the \rmat calculations are discussed in \secab\ref{sec:rmat}. The phase 
	shifts and cross sections are analyzed in \secab\ref{sec:phscs}.
	\section{Model Potential of L\lowercase{i}$^+$}
	\label{sec:modelpot}
	In the calculations discussed below, the target atom is represented by its valence electron in 
	the presence of the spherically symmetric potential $V_{l_1}(r)$ that models the closed-shell 
	core of Li$^+$. This model potential is constructed individually for every angular momentum 
	$l_1$ of the valence electron. It is optimized in such way that the energies of the low-lying 
	bound states supported by $V_{l_1}(r)$ coincide with the energies of the ground and low excited 
	states of the lithium atom.
	
	The form of $V_{l_1}(r)$ used in this work is
	\begin{eqnarray}
		V_{l_1}(r)=-\frac{1+2\exp(-a_{l_1}r)+b_{l_1}r\exp(-c_{l_1}r)}{r}\nonumber\\
		-\frac{\alpha_d}{2r^4}W_6(\rho_{l_1},r),
		\label{eq:modelpot}
	\end{eqnarray}
	where $a_{l_1}$, $b_{l_1}$, $c_{l_1}$ and $\rho_{l_1}$ are the parameters to be optimized, $\alpha_d=0.189$~a.u. is the 
	polarizability of the Li$^+$ core \cite{Pouchan1984} and
	\begin{equation}
		W_n(r_c,r)=1-\exp\left[-(r/r_c)^n\right]
	\end{equation}
	is the cut-off function regularizing the potential at the origin. \eqab\eqref{eq:modelpot} is a generalization of the 
	potential employed by \citet{Pan1996} that is $l_1$-independent and the polarization part of the potential vanishes less 
	rapidly with decreasing value of $r$ than in $V_{l_1}(r)$ constructed in this work. Very similar $l_1$-dependent model 
	potential was developed to represent the Li$^+$ core by \citet{Marinescu1994} in their research of the dispersion 
	coefficients for the alkali-metal dimers. Generally, in the research of the interactions between electrons in the continuum 
	and neutral atoms or positive ions, the cationic cores have been very successfully modeled by this form of the potential (see 
	\refsab\cite{Greene1990,Pan1996,Marinescu1994} as well as \cite{Aymar1996} and references therein).
	
	The set of parameters $a_{l_1}$, $b_{l_1}$, $c_{l_1}$ and $\rho_{l_1}$ was optimized using the 
	non-linear least squares method independently for $l_1=0,1,2$. The accurate theoretical 
	\cite{Frolov20141} and experimental \cite{Moore1971} energies of five lowest states with 
	respect to the ionization threshold for every $l_1=0\dots2$ were taken as the data to be 
	matched by the model. In every iteration, it was necessary to diagonalize the one-particle 
	Hamiltonian operator $\hat{H}_{l_1}=\hat{K}+V_{l_1}(r)$ where $\hat{K}$ is the operator of the kinetic energy. 
	The match of the obtained eigenenergies $\varepsilon_{nl}$ with the experimental data then 
	determined the adjustments of $V_{l_1}(r)$ in the next iteration. Note that the index $n$ plays a role of the principal 
	quantum number as known in the atomic physics. In the calculations presented 
	here, $\hat{H}_{l_1}$ was represented by the radial basis set consisting of 2000 \bspls 
	\cite{Bachau2001} that spanned the sphere with radius 240~a.u. This size of the sphere was 
	chosen with respect to the fact that the classical turning point of the highest fitted bound 
	state is at $\sim95$~a.u. Sufficient radial interval beyond this limit allowed for the accurate 
	exponential decrease of the wave function and eliminated the artifacts of the finite box.
	\begin{table}
		\caption{Optimized values of the parameters of $V_{l_1}(r)$ in \eqab\eqref{eq:modelpot} for Li$^+$.}
		\label{tab:potparams}
		\begin{ruledtabular}
			\begin{tabular}{cdddd}
				&\multicolumn{1}{c}{$l_1=0$}&\multicolumn{1}{c}{$l_1=1$}&\multicolumn{1}{c}{$l_1=2$}\\
				\hline
				$a_{l_1}$&10.655&2.734&25.915\\
				$b_{l_1}$&3.397&46.621&7.562\\
				$c_{l_1}$&2.821&10.493&18.200\\
				$\rho_{l_1}$&0.375&1.294&1.258
			\end{tabular}
		\end{ruledtabular}
	\end{table}
	The values of the parameters optimized to represent Li$^+$ in the \rmat calculations discussed 
	below are listed in \tabab\ref{tab:potparams}. The model potential $V_{l_1}(r)$ with these 
	parameters yielded less than 1~meV deviation of the  calculated energy levels from lowest five 
	experimental values \cite{Moore1971} for every $l_1\leq2$.
	
	The best match between the experimental energies of the $s$-states and the spectrum of the 
	$\hat{H}_0$ was achieved when the lowest eigenvalue $\varepsilon_{10}$ was omitted from the 
	optimization of $V_{l_1}(r)$ and the second eigenenergy $\varepsilon_{20}$ was compared with 
	the ground state of the lithium atom. This is related to the fact that the 1$s$ orbital in 
	lithium is doubly occupied by the core electrons and the lowest $s$-orbital available for the 
	valence electron is $n=2$ that possesses one node. As a result, $\hat{H}_0$ supports one very 
	deeply bound core-like non-physical state with the energy $\varepsilon_{10}=-2.0069$~a.u. with 
	respect to the ionization threshold. This orbital is very compact, its classical turning point 
	is located at $\sim0.75$~a.u. The eigenenergies $\varepsilon_{n0}$ above this state very 
	accurately correspond to the experimental energy levels of the lithium atom \cite{Moore1971}.
	
	For $l_1>2$, the energies of the lithium bound states are so close to the corresponding levels of the hydrogen atom that with very good approximation the Coulomb potential $-1/r$ can be taken instead of $V_{l_1}(r)$.
	
\begin{figure}
		\includegraphics[]{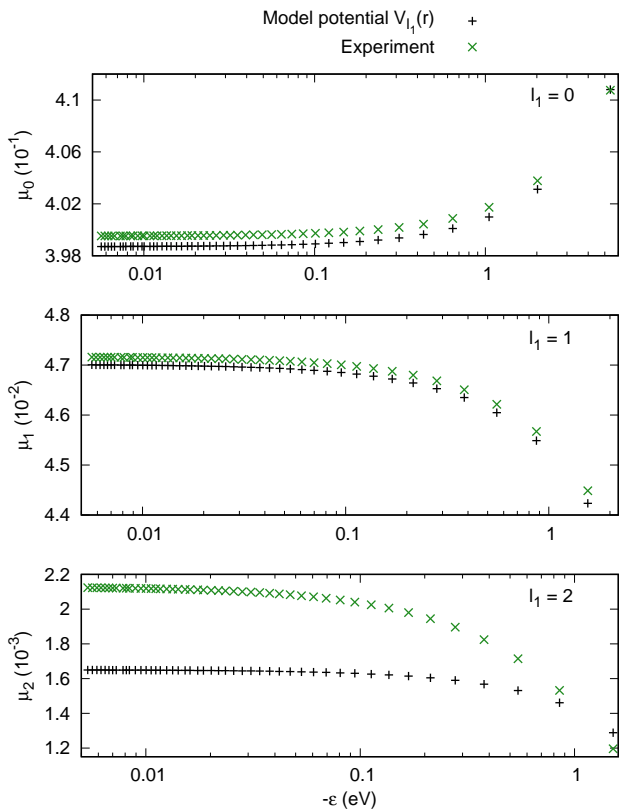}
		\caption{Quantum defects $\mu_{l_1}(\varepsilon)$ of Li as functions of the negative energy $\varepsilon$ (logarithmic scale) taken with respect to the ionization threshold. The top, center and bottom panel shows the results for $ l_1=0,1,2$, respectively. The values obtained using the model potential $V_{l_1}(r)$ in \eqab\eqref{eq:modelpot} and parameters from \tabab\ref{tab:potparams} ($+$) are compared with the experimental results published by \citet{Goy1986} for $l_1=0,1$ and by \citet{Lorenzen1983} for $l_1=2$ ($\times$).}
		\label{fig:qds}
	\end{figure}

	Although it is not the main objective of this work, it is interesting to mention that 
	$V_{l_1}(r)$ also yields accurate energies of the excited states higher than those to which 
	$V_{l_1}(r)$ was optimized. \figab\ref{fig:qds} shows the quantum defects 
	$\mu_{l_1}(\varepsilon)$ calculated for $V_{l_1}(r)$ \cite{Seaton1966}. Its good correspondence 
	to the experimental results \cite{Goy1986,Lorenzen1983} (note the order-of-magnitude decrease 
	of $\mu_{l_1}(\varepsilon)$ with increasing value of $l_1$) implies that $V_{l_1}(r)$ also 
	correctly models the $s$-, $p$- and $d$- Rydberg states of the Li atom.

	The aim of the extensive radial basis set utilized in the optimization discussed above is to eliminate the effects of the finite basis set as much as possible and to provide the model potential that is independent of the basis set. Note that this approach is different from the method frequently used in quantum chemistry to represent the atomic cores by the potentials. In such calculations, mainly based on the Gaussian basis sets, the parameters of the potential are optimized for one specific basis set that becomes part of the model along with the optimized potential (see \cite{Guerout2010} and references therein).
	
	In the calculations of the electron collisions with lithium at low energies, 
	\citet{Norcross1971} also used a model potential to represent the Li$^+$ core. The parameters 
	of the scaled Thomas-Fermi potential \cite{Stewart1965} with additional polarization term were 
	optimized to accurately reproduce the energies of two lowest eigenstates of the lithium atom. 
	The Thomas-Fermi potential was also utilized by \citet{Moores1986} to calculate the 1$s$ core 
	wave functions of Li$^+$. The valence orbitals of the neutral lithium were obtained from the 
	\electron-Li$^+$ scattering calculations. On the other hand, \citet{Burke1969}, in their work, represented 
	the lithium atom by the Hartree-Fock wave function and constrained the 1$s$ orbital to be 
	doubly 
	occupied in all the terms considered in the following CC expansion of the scattering wave 
	function.
	
	In the two-electron calculations, the approximation of the noble-gas-like core by the model 
	potential $V_{l_1}(r)$ can be corrected by including the dielectronic term introduced by 
	\citet{Chisholm1964} in the two-electron Hamiltonian:
	\begin{eqnarray}
		V_{\text{diel}}(\rho_c, r_1,r_2, \theta_{1,2})=\nonumber\\
		-\frac{\alpha_d}{r_1^2r_2^2}\left[W_6(\rho_c,r_1)W_6(\rho_c,r_2)\right]^{1/2}P_1(\cos\theta_{12})\nonumber\\
		-\frac{\alpha_q}{r_1^3r_2^3}\left[W_{10}(\rho_c,r_1)W_{10}(\rho_c,r_2)\right]^{1/2}P_2(\cos\theta_{12}),
		\label{eq:dielec}
	\end{eqnarray}
	where $r_1$ and $r_2$ are the radial coordinates of the first and second electron, respectively, $\theta_{1,2}$ is the angle 
	between their position vectors, $P_n(x)$ is the $n$th Legendre polynomial, $\rho_c$ is the cut-off parameter and $\alpha_q$ 
	is the quadrupole polarizability of the core. This term describes the interaction between the valence and scattering electrons via the dipole and quadrupole moments induced on the core. Although this correction becomes more important for the heavier 
	alkali metals, it was included in the \rmat calculations presented in this work with $\alpha_q=0.037$~a.u. 
	\cite{Maroulis1986}. The value of the cut-off parameter $\rho_c=4.03$~a.u. was chosen in such way that the electron affinity 
	of Li$^-$ obtained by the diagonalization of the two-electron Hamiltonian discussed in \secab\ref{sec:rmat} including the 
	correction \eqref{eq:dielec} coincides with the accurate experimental value 0.617~eV \cite{Dellwo1992}. The electron affinity 
	calculated using the value of $\rho_c$ mentioned above is 0.620~eV.
	\section{\rmat Calculations}
	\label{sec:rmat}
	The scattering calculations discussed below were performed using the \rmat computer program by 
	\citet{Tarana2016} originally designed to calculate the electronic states of the long-rang 
	Rydberg molecules. The notation introduced in \refab\cite{Tarana2016} was adopted in this section. 
	The reader is also referred there for the definitions of the open and closed one-particle wave 
	functions and two-electron configurations (see also \refab\cite{Aymar1996}). Only the 
	inner-region part of the program by \citet{Tarana2016} was used in this work. The outer-region 
	code was developed independently and it consists of the propagation of the \rmat in the 
	long-range potentials of the target \cite{Baluja1982,Pan1996} as well as of the construction of 
	the $K$-matrix, $T$-matrix and calculation of the phase shifts and cross sections 
	\cite{Zatsarinny,Taylor2006book,Nesbet1980book}.
	
	The radius of the \rmat sphere was set to $r_0=120$~a.u. This allows for equally accurate 
	treatment of both short-range and long-range interactions between the target atom and the 
	incident electron. The long-range effects become particularly important at the scattering 
	energies near the \twos threshold. This is the energy range from which the accurate values of 
	the MERT parameters can be obtained. Since the full interaction of the electrons with each 
	other as well as with the Li$^+$ core is considered in the inner region, the treatment of the 
	long-range effects inside this relatively large sphere is not restricted only to the potential 
	due to the static dipole polarizability of the target. It also includes the effects of the 
	higher multipoles of the ground and excited states. Similarly large spheres were used by 
	\citet{Pan1996} in their \rmat calculations of the photodetachment of Li$^-$.
	
	The set of 154 radial \bspls of the 6th order was used inside the \rmat sphere to represent the 
	closed and open single-particle wave functions \cite{Tarana2016}. For every one-electron 
	angular momentum $l\leq7$, 25 lowest closed orbitals were included in the closed part of the 
	ansatz for the two-electron wave function. All possible excitations involving these orbitals 
	were included in the construction of the corresponding configuration interaction (CI) 
	Hamiltonian matrix $\underline{H}'$. Lowest 6, 6, 5, 3,  and 1 closed orbitals among the $s$-, 
	$p$-, $d$-, $f$- 
	and $g$-states, respectively, were included in the open part of the two-electron wave function 
	ansatz \cite{Tarana2016} as the scattering channels. This extensive basis set ensures very 
	accurate treatment of all the correlation and polarization effects. Our tests showed that 
	further augmentation 
	of the basis set has negligible impact on the calculated scattering quantities.
	
	The CI matrix $\underline{H}'$ representing the two-electron Hamiltonian in the inner region 
	(including the dielectronic term \eqref{eq:dielec} and Bloch operator 
	\cite{Aymar1996,tennyson-rev,Zatsarinny,Tarana2016}) was diagonalized and using the eigenvalues 
	(\rmat poles) $E_k$, the \rmat was calculated as
	\begin{equation}
		R_{\bar{j}\bar{j}'}(E)=\frac{1}{2}\sum_k\frac{w_{\bar{j}k}w_{\bar{j}'k}}{E_k-E},
		\label{eq:rmat}
	\end{equation}
	where $E$ is the total energy of the \electron-Li system and $w_{\bar{j}k}$ are the surface 
	amplitudes -- projections of the $k$th eigenstate of $\underline{H}'$ on the target state $n$ 
	with the angular momentum $l_1$ and on the partial wave $l_2$ of the scattered electron 
	\cite{Aymar1996,tennyson-rev,Zatsarinny,Tarana2016}. The 
	multi-index $\bar{j}=\{n,l_1,l_2\}$ denotes the scattering channel.

	Since the total angular momentum $L$, total spin $S$ and total parity $P=(-1)^{l_1+l_2}$ of the \electron-Li system are good quantum numbers, the scattering calculation can be performed independently for each $LSP$ symmetry and the cross sections calculated in this way can be summed to obtain the results that can be compared with the experiments.
	
	It is worth mentioning at this point that the spectrum of $\underline{H}'$ includes a set of 
	non-physically low \rmat poles $E_k$. This is an artifact of the very low-lying compact orbital 
	with energy $\varepsilon_{10}$ discussed in \secab\ref{sec:modelpot}. In the eigenstates 
	corresponding to these low-lying \rmat poles, the configurations where the compact 1$s$-like 
	orbital is singly- or doubly-occupied are dominant and not strongly coupled to the 
	configurations involving the higher valence orbitals. Since, in addition, this core-like target 
	state was not included in the CC expansion as the scattering channel, there are no 
	surface amplitudes $w_{\bar{j}k}$ associated with it. As a result, these non-physically 
	low-lying eigenstates of $\underline{H}'$ do not appear in the pole-expansion \eqref{eq:rmat} 
	of the \rmat and they do not affect the results of the scattering calculations.
	
	This work is dealing with the kinetic energies of the incident electron $\epsilon\leq3.3$~eV 
	where only the \twos and \twop channels are open. The \twop channel opens at the energy 
	$\epsilon=1.848$~eV above the \twos threshold and the threshold of the \threes channel is 
	located at 
	$\epsilon=3.373$~eV. Although all the remaining higher channels included in the scattering 
	calculations are closed, their presence ensures the accurate treatment of the long-range 
	\electron-Li interaction in the outer region represented by the transition dipole moments 
	coupling the target states \cite{Pan1996,tennyson-rev}. In spite of the large \rmat box, the 
	propagation of the \rmat \cite{Baluja1982} in the long-range tail of the lithium potential to 
	the distance 2700~a.u. from the center was necessary to obtain converged phase shifts at 
	energies below 1~meV suitable for the calculation of the MERT parameters.
	
	One way to assess the representation of the \electron-Li interaction by the dipole potentials coupling the target states outside the \rmat sphere  is to calculate the static dipole polarizabilities $\alpha_{nl_1}$ of the ground and excited states of the target from the dipole moments used in the present \rmat propagation \cite{Friedrichbook}:
	\begin{equation}
		\alpha_{nl_1}=2\sum_{\substack{l_1'=l_1\pm1\\n'\neq 
		n}}\frac{\left|\braket{\psi_{n'l_1'}|z|\psi_{nl_1}}\right|^2}{\varepsilon_{n'l_1'}-\varepsilon_{nl_1}},
		\label{eq:polari}
	\end{equation}
	where $\ket{\psi_{n'l_1'}}$ and $\ket{\psi_{nl}}$ are the eigenstates of the target and the 
	matrix element in the numerator of \eqab\eqref{eq:polari} is the $z$-component of the 
	corresponding transition dipole moment. This equation yields polarizability of the lithium 
	atom in the ground state $\alpha_{20}=163.7$~a.u. that is in excellent agreement with 
	previously published theoretical and experimental values varying between 163.74~a.u. and 164.19~a.u. (see 
	\refab\cite{Puchalski2011} and references therein). The contribution from the $2s\to2p$ 
	term to the series \eqref{eq:polari} represents more than 99\% of the calculated value and all 
	the higher terms are its small corrections.
	
	For the excited state \twop, \eqab\eqref{eq:polari} yields the static dipole polarizability 
	$\alpha_{21}=109.35$~a.u. that is lower than previously published theoretical and experimental 
	values varying between  125.2~a.u. and 135.7~a.u. (see \cite{Tang2009} and references therein). 
	This suggests that including only the excited target states from the \rmat channel space is not sufficient to accurately 
	treat the polarizability $\alpha_{21}$ in \eqab\eqref{eq:polari}. However, our test 
	scattering calculations (not presented in this paper) showed that including more target states 
	in the expansion of the scattering wave function has negligible influence on the calculated 
	phase shifts and cross sections.
	
	The \rmat propagated to the distance 2700~a.u. from the center is used to match the linear combination of the regular and 
	irregular free-particle solutions of the Schr\"odinger equation in the open channels \cite{Zatsarinny}. This yields the 
	$K$-matrix and its subsequent diagonalization provides the scattering phase shifts. The scattering amplitudes and cross 
	sections are then calculated using the standard methods of the multi-channel scattering theory \cite{Taylor2006book}. It is 
	not necessary to consider the closed channels in the matching of the free-particle solutions and to perform any elimination 
	of the closed channels \cite{Aymar1996}. In this case, the energy range around the threshold where the closed channels 
	influence the results, is negligible.
	\section{Results}
	\label{sec:phscs}
	\subsection{Phase shifts}
	In this section, the dependence of the phase shifts $\delta_{l_2}^{2S+1}(\epsilon)$ on the 
	kinetic energy of the incident electron $\epsilon=k^2/2$ is presented as follows: Below the 
	\twop threshold, $\delta_{l_2}^{2S+1}(\epsilon)$ is a single phase shift corresponding to the 
	partial wave $l_2$ of the scattered electron according to the total $LSP$ symmetry of the 
	\electron-Li system. Above the 2$p$ threshold, $\delta_{l_2}^{2S+1}(\epsilon)$ is the sum of all 
	the eigenphases obtained by the diagonalization of the multi-channel $K$-matrix. The 
	superscript $2S+1$ denotes the multiplicity of the \electron-Li system. Since the parity $P=(-1)^{l_1+l_2}$ and since the 
	electron 
	collisions with lithium in its excited states are not a subject of this work, it is sufficient 
	to deal only with those $LSP$ symmetries where $P=(-1)^L$ as only in these the colliding 
	electron is coupled with the ground state of the target. Therefore, it is not necessary to use 
	parity index to denote the phase shifts in different $LSP$ symmetries.
	
\begin{figure}
		\includegraphics{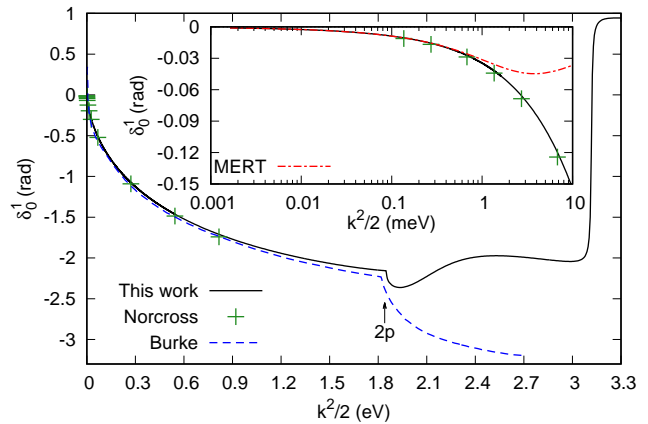}
		\caption{\seone phase shift as a function of the kinetic energy of the scattered electron. The results calculated in this 
		work (full black line) are compared with previously published data by \citet{Norcross1971} (+) and \citet{Burke1969} 
		(blue dashed line). The inset shows the detail of the phase shifts in the range 0.01 -- 10~meV using the logarithmic 
		energy scale. The red dash-dotted line is the MERT fit of the current results ($A_1=2.815$).}
		\label{fig:phs1se}
	\end{figure}

	The \seone phase shift calculated using the \rmat method is plotted in \figab\ref{fig:phs1se}. Its comparison with the 
	results calculated by \citet{Norcross1971} shows an excellent agreement  for the energy range between 0.1~meV and 0.8~eV.
	The rapid decrease at very low scattering energies is the consequence of the \seone bound state of Li$^-$. The low-energy 
	$s$-wave phase shifts calculated in this work can be parametrized by the MERT \cite{OMalley1961}
	\begin{eqnarray}
		\frac{k}{\tan\delta_0^{2S+1}(k)}=-\frac{1}{A_{2S+1}}+\frac{\pi\beta^2}{3A_{2S+1}^2}k\nonumber\\
		+\frac{4\beta^2}{3A_{2S+1}}k^2\ln\left(\frac{\beta k}{4}\right)
		+\left[\frac{r_{0,2S+1}}{2}+\frac{\pi\beta}{3}\right.\nonumber\\
		\left.+\frac{20\beta^2}{9A_{2S+1}}-\frac{8\beta^2}{3A_{2S+1}}\gamma-\frac{\pi\beta^3}{3A_{2S+1}^2}-\frac{\pi^2\beta^4}{9A_{2S+1}^3}\right]k^2,
	\end{eqnarray}
	where $A_{2S+1}$ is the scattering length, $r_{0,2S+1}$ denotes the 
	effective range, $\beta^2=\alpha_{20}$ and $\gamma=\Gamma'(3/2)/\Gamma(3/2)=0.0365$ is the logarithmic derivative of the 
	Gamma function.
	This fit yields the scattering length $A_1=2.815$ and corresponding effective range $r_{01}=634.808$. While the optimized 
	value of 
	$r_{01}$ is sensitive to the energy interval taken for the non-linear fit, the scattering length $A_1$ does not considerably 
	change.	Moreover, if the ground-state polarizability $\alpha_{20}$ is treated as a fitting parameter, its value from the 
	inner-region calculations is reconstructed and the scattering length $A_1$ remains unchanged.
	
	The \seone scattering length obtained from the \rmat calculations discussed here is slightly lower than the 
	value 3.04 reported by \citet{Norcross1971}. Possible reason for this subtle difference can lay in the fact that the \rmat 
	propagation utilized in this work allows for stable evaluation of $\delta_0^1(\epsilon)$ at lower energies than those 
	considered by \citet{Norcross1971}. As a result, the MERT parametrization performed at lower collision energies can also 
	yield slightly different value of $A_1$. As can be seen in the inset of \figab\ref{fig:phs1se}, the MERT parametrization 
	\cite{OMalley1961} of the \seone phase shift considerably deviates from the \rmat results at scattering energies above 1~meV.
	
	Below the \twop excitation threshold, the \rmat calculations also show an agreement with the 
	results obtained by \citet{Burke1969}. The discrepancy increases at the scattering energies above the \twop threshold due to 
	the 
	terms in the CC expansion involving the target orbitals with higher energies and angular momenta that are not 
	included in \cite{Burke1969}. However, they are necessary for accurate representation of the interaction between the incident 
	electron and the valence electron of the target. This issue of the truncated CC expansion in \cite{Burke1969} is 
	not specific only for the \seone scattering but it is common for all the $LSP$ symmetries.
	
	Another structure in the calculated \seone phase shifts that can be seen in \figab\ref{fig:phs1se} is the narrow resonance 
	below the \threes threshold. It can be very accurately fitted to the Breit-Wigner formula \cite{Taylor2006book}
	\begin{subequations}
		\begin{equation}
			\delta(\epsilon)=\delta_{\text{bg}}(\epsilon)+\delta_{\text{res}}(\epsilon),
		\end{equation}
		where $\delta_{\text{bg}}(\epsilon)$ is the background phase shift slowly varying with the energy,
		\begin{equation}
			\delta_{\text{res}}(\epsilon)=\arcsin\left[\frac{\Gamma/2}{\sqrt{\left(\epsilon-E_r\right)^2+(\Gamma/2)^2}}\right],
			\label{eq:bwres}
		\end{equation}
		\label{eq:bwphs}
	\end{subequations}
	$E_r$ is the position of the resonance and $\Gamma$ is its width. Since \eqsab\eqref{eq:bwphs} can be applied to the 
	resonances in any $LSP$ symmetry, the indices $l$ and $2S+1$ are omitted from the notation of the resonant and background 
	phase shifts. When $ \delta_{\text{bg}}(\epsilon)$ is assumed to be slowly varying function of the energy
	\begin{equation}
		\delta_{\text{bg}}(\epsilon)=P_0+P_1\epsilon,
		\label{eq:linbgphs}
	\end{equation}
	the fit yields $E_r=3.117$~eV and $\Gamma=11$~meV. The optimized values of $P_0$ and $P_1$ are listed in 
	\tabab\ref{tab:phsparams}.
	\begin{table*}
		\caption{Summary of the scattering lengths $A$ and parameters $B$ in \eqab\eqref{eq:pmert} calculated for different $LSP$ 
		symmetries and their comparison with values previously published by \citet{Norcross1971}. Parameters of the resonances 
		are obtained by fitting the calculated phase shifts to \eqab\eqref{eq:bwphs}.}
		\label{tab:phsparams}
		\begin{ruledtabular}
			\begin{tabular}{cccccccc}
				$LSP$&\multicolumn{2}{c}{MERT}&\multicolumn{5}{c}{Resonances}\\
				\cline{2-3}\cline{4-8}
				symmetry&\rmat&\citet{Norcross1971}&$E_r$ (eV)&$\Gamma$ (meV)&$P_0$&$P_1$&$P_2$\\
				\hline
				\seone&$A_1=2.815$, $r_{01} = 634.808$&$A_1=3.04$&3.117&11&1.932&-0.292&\\
				\sethree&$A_3=-7.46$, $r_{03}=4.89$&$A_3=-7.12$\\
				\poone&$B_1=-1.85$&$B_1=-1.69$\\
				\hline
				\pothree&$B_3=0.236$&$B_3=1.992$&0.062&68&0.504&-2.582&2.423\\
				&&&3.285&45&0.107&0.087\\
			\end{tabular}
		\end{ruledtabular}
	\end{table*}
	The analysis of the CI configurations contributing to the resonant wave function reveled that this resonance has Feshbach 
	character with the dominant configuration $3s^2$. To our best knowledge, this resonance has not been reported in any 
	previously published papers dealing with the \electron-Li collisions. Due to its symmetry, this resonance is expected to 
	appear in the two-photon detachment spectrum of Li$^-$ at the photon energy $\sim0.072$~a.u.  Similar spectrum was calculated 
	by \citet{Glass1998}. However, the highest photon energy considered in that work was 0.05~a.u.
	\begin{figure}
		\includegraphics{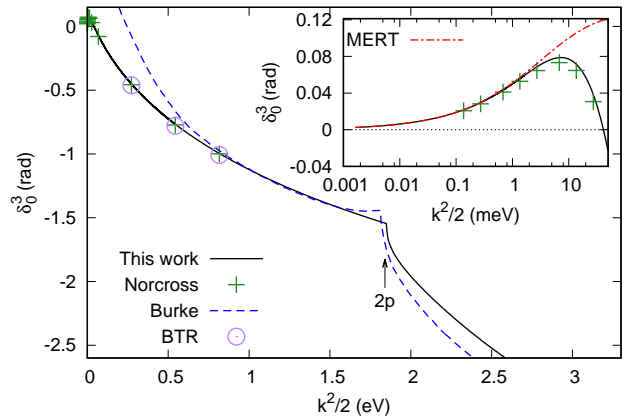}
		\caption{\sethree phase shift as a function of the kinetic energy of the scattered electron. The results calculated in 
		this work (full black line) are compared with previously published data by \citet{Norcross1971} (+) and by 
		\citet{Burke1969} 
		(blue dashed line). The circles ($\circ$) represent the results of \citet{Norcross1971} obtained after the orthogonality 
		correction to Burke's procedure. The inset shows the detail of the phase shifts in the range 0.01 -- 14~meV on 
		logarithmic energy scale. The red dash-dotted line is the MERT fit of the current results ($A_3=-7.46$).}
		\label{fig:phs3se}
	\end{figure}
	
	\figab\ref{fig:phs3se} shows the comparison of the \sethree phase shifts calculated using the \rmat method with previously 
	published results. Like in the case of singlet discussed above, the data obtained in this work are in very good agreement 
	with those calculated by \citet{Norcross1971}. As a result, the energy at which $\delta_0^3(\epsilon)$ changes from positive 
	to negative (that corresponds to the Ramsauer-Townsend minimum in the cross section) calculated  in this work also matches 
	very well with the value obtained by \citet{Norcross1971}. Fitting of the calculated phase shifts to MERT \cite{OMalley1961} 
	yields the scattering length $A_3=-7.46$ and the effective range $r_{03}=4.89$ that is consistent with the previously 
	published value $A_3=-7.12$ \cite{Norcross1971}. As can be seen in the inset of \figab\ref{fig:phs3se}, the MERT 
	parametrization of $\delta_0^3(\epsilon)$ is valid at energies below 5~meV. The steep increase of the $s$-wave phase shift at 
	very low energies followed by the rapid drop is characteristic for the scattering systems that possess the virtual state 
	\cite{Taylor2006book} (represented by a pole of the $S$-matrix on the negative imaginary axis in the complex momentum plane) 
	with sufficiently small energy.
	
	The reason behind the deviation of the \sethree phase shifts calculated by \citet{Burke1969} from the results presented in 
	this work and those obtained by \citet{Norcross1971} is that the CC method, as formulated by \citet{Burke1969}, does not 
	guarantee that the continuum wave functions of the colliding electron are orthogonal to the Hartree-Fock orbitals of the 
	target \cite{Norcross1969}. When \citet{Norcross1969,Norcross1971} introduced this orthogonality into the CC equations 
	formulated by \citet{Burke1969} as the additional constraint (circles denoted as BTR in \figab\ref{fig:phs3se}), the phase 
	shifts calculated in this way became consistent with those obtained using other approaches discussed in this work. This 
	orthogonality issue is common to all the $LSP$ symmetries discussed here, except the \seone scattering \cite{Norcross1969}. 
	It is another limitation of the computational method used in \refab\cite{Burke1969} in addition to the truncation of 
	the CC expansion mentioned above.
	
\begin{figure}
		\includegraphics{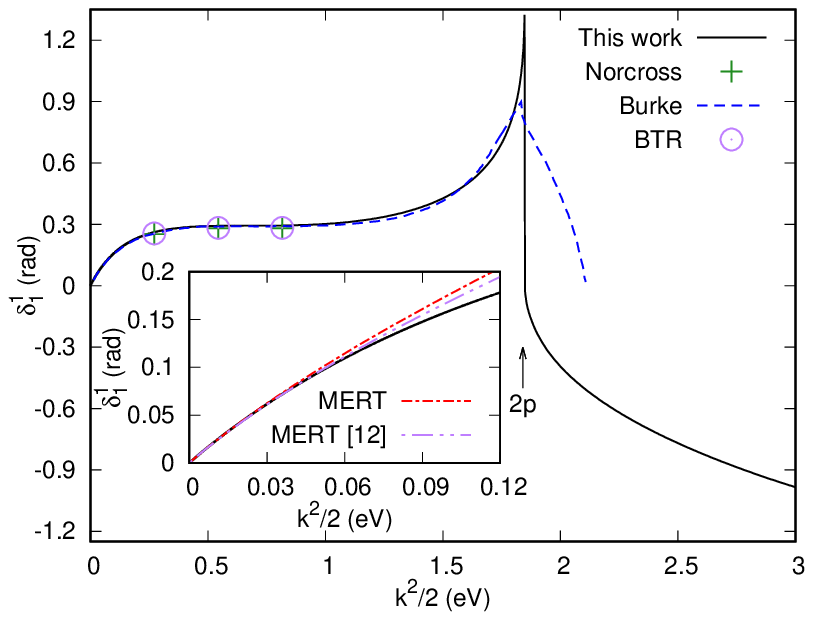}
		\caption{\poone phase shift as a function of the kinetic energy of the colliding electron.
		The data are displayed with the same symbols as in \figab\ref{fig:phs3se}.
		The inset shows the detail of the phase shifts in the range below 120~meV. The red 
		dash-dotted line corresponds to the MERT fit of the current results ($B_1=-1.85$), the violet dash-dot-dotted line is the 
		MERT with $B_1=-1.69$ taken from \refab\cite{Norcross1971}.}
		\label{fig:phs1po}
	\end{figure}

	The \poone phase shifts calculated using the \rmat method are in excellent agreement with the results of \citet{Norcross1971} at all three energy points where the latter are provided (see \figab\ref{fig:phs1po}).
	In \refab\cite{Norcross1971}, these were extrapolated towards the low energies near the threshold using the $p$-wave 
	MERT \cite{OMalley1961}, as
	\begin{equation}
		\delta_1^{2S+1}(k)=\frac{\pi\beta^2 k^2}{15}+\frac{\beta^3k^3}{9B_{2S+1}},
		\label{eq:pmert}
	\end{equation}
	with the value of the fitting parameter $B_1=-1.69$. The parametrization of our \rmat results at the scattering 
	energies below 10~meV using \eqab\eqref{eq:pmert} yields similar value $B_1=-1.85$. As can be seen in the inset of 
	\figab\ref{fig:phs1po}, this parametrization fits the \abini results for energies below 50~meV. In spite of the issue with 
	the orthogonality of the continuum wave functions and target orbitals discussed above, at energies below the \twop threshold, 
	the phase shifts obtained by \citet{Burke1969} are in encouraging agreement with our \rmat calculations and the orthogonality 
	correction \cite{Norcross1971} does not considerably change the results. Rigorously, \eqab\eqref{eq:pmert} above should 
	contain an additional term $\sim k^3$ not involving $\beta$ \cite{OMalley1961}. Its presence is dictated by the Wigner 
	threshold law for the short-range interactions. Since only the whole coefficients in front of $k^3$ can be fitted and since 
	$\beta$ is constant, we used the form \eqref{eq:pmert} in order to have like-to-like comparison with the MERT fit by 
	\citet{Norcross1971}.

\begin{figure}
		\includegraphics{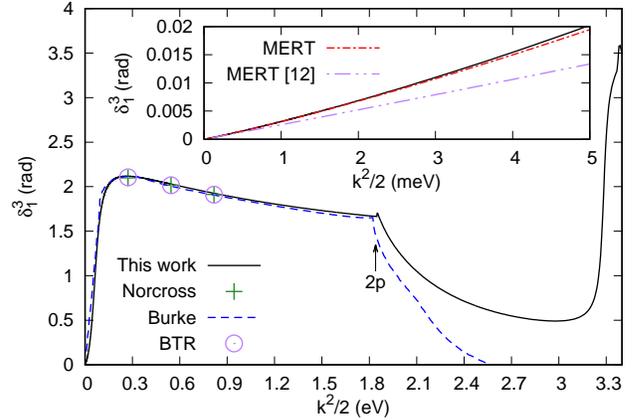}
		\caption{\pothree phase shift as a function of the kinetic energy of the scattered electron. 
		The data are displayed with the same symbols and lines as in \figab\ref{fig:phs3se}.
		The inset shows the detail of the phase shifts in the range below 5~meV. The red 
		dash-dotted line corresponds to the MERT fit of the current results ($B_3=0.236$), the violet dash-dot-dotted line 
		represents the 
		MERT with $B_3=1.992$ taken from \refab\cite{Norcross1971}.}
		\label{fig:phs3po}
	\end{figure}

	\citet{Norcross1971} also performed the low-energy $p$-wave MERT extrapolation of the \pothree phase shift calculated at the 
	same three energy points between 0.2~eV and 0.8~eV as in the \poone symmetry and obtained the value of the fitting parameter 
	$B_3=1.992$. However, the adequacy of this extrapolation can be called into question as the there is a \pothree resonance 
	below 100~meV 
	that changes the threshold behavior of the phase shift and the MERT expansion may not be valid above this energy (see 
	\figab\ref{fig:phs3po}).
	The value optimized by fitting our \rmat results to \eqab\eqref{eq:pmert} at energies sufficiently below this resonance is 
	considerably smaller, $B_3=0.236$. As can be seen in the inset of \figab\ref{fig:phs3po}, this parametrization is more 
	consistent with the \abini results than the parametrization in \refab\cite{Norcross1971}.
	
	The fit of the low-energy \pothree resonance shown in \figab\ref{fig:phs3po} is disturbed by the threshold effects. 
	The phase shift in the corresponding energy interval can be accurately fitted to \eqsab\eqref{eq:bwphs} assuming the near-threshold behavior of the background phase shift consistent with \eqab\eqref{eq:pmert}
	\begin{equation}
		\delta_{\text{bg}}(\epsilon)=P_0+P_1\epsilon+P_2\epsilon^{3/2},
	\end{equation}
	where the value of $P_0=-\delta_{\text{res}}(0)$ is chosen so that $\lim_{\epsilon\to0}\delta(\epsilon)=0$, $P_1$ and $P_2$ 
	are the fitting variables. This optimization yields $E_r=62$~meV and 
	$\Gamma=68$~meV. The optimized values of $P_0$, $P_1$ and $P_2$ are listed in \tabab\ref{tab:phsparams}. Generally, the 
	interaction of 
	the resonance with the threshold can be more complicated than present ad hoc assumption that the phase shift is simply a sum 
	of 
	the Breit-Wigner formula \eqref{eq:bwres} and $p$-wave MERT expression \eqref{eq:pmert}. The narrower the fitted resonance 
	is, the more 
	accurate values of $E_r$ and $\Gamma$ this approach yields since the rapid increase of $\delta_1^3(\epsilon)$ occurs at 
	smaller energy scale than the steady change due to the threshold behavior. Therefore, this parametrization provides a 
	numerically accurate model of the low-energy behavior of the \pothree phase shift suitable for the construction of related 
	zero-range potentials \cite{Omont1977,Hamilton2002}.
	
	Another \pothree resonance appears just below the \threes excitation threshold (see \figab\ref{fig:phs3po}) and, assuming 
	that 
	$\delta_{\text{bg}}(\epsilon)$ is linear (\eqab\eqref{eq:linbgphs}), it can be parametrized using the Breit-Wigner formula 
	\eqref{eq:bwphs} where $E_r=3.285$~eV and $\Gamma=45$~meV.
	
\begin{figure}
		\includegraphics{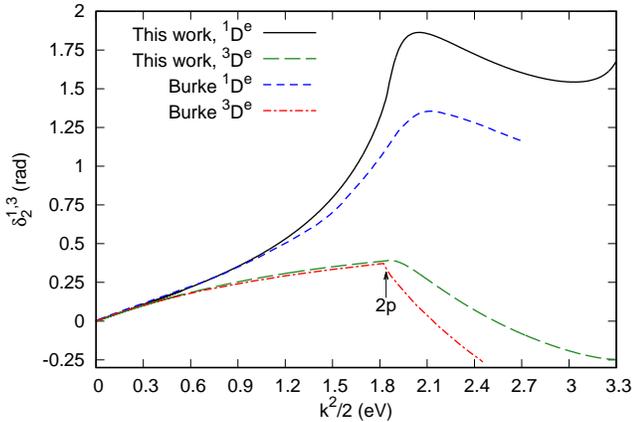}
		\caption{\deone and \dethree phase shifts calculated using the \rmat method as functions of the kinetic energy of the colliding electron (solid black line and long-dashed green line, respectively) are compared with the \deone and \dethree results calculated by \citet{Burke1969} (short-dashed blue line and dotted-dashed red line, respectively).}
		\label{fig:phs13de}
	\end{figure}

	The \deone and \dethree phase shifts calculated in this study are plotted in \figab\ref{fig:phs13de} along with the results 
	obtained by \citet{Burke1969}.
	While the triplet phase shifts agree very well below the \twop threshold, the singlet results of \citet{Burke1969} are, for 
	energies above 1.2~eV, lower than those obtained from the \rmat calculations presented here. The phase shift steeply raises 
	above this energy reaching the highest value above the \twop threshold. Likewise to the low-energy \pothree resonance 
	discussed 
	in 
	the text above, the fit of this resonance by the  Breit-Wigner formula \eqref{eq:bwphs} is complicated by its interaction 
	with the threshold. In this case, however, the resonant structure in the phase shifts reaches the energy regions both below 
	and above the 2p threshold. Therefore, $\delta_{\text{bg}}(\epsilon)$ has different energy dependence below and above the 
	\twop threshold. Since this resonance is relatively broad compared to the other resonances discussed above, the steep 
	increase of 
	the phase shift cannot be attributed only to $\delta_{\text{res}}(\epsilon)$. This makes the values of $E_r$ and $\Gamma$ 
	very sensitive to the form of $\delta_{\text{bg}}(\epsilon)$ and to the energy range taken for the fit.
	Note that this resonance also appears in the two-photon detachment spectrum of Li$^-$ calculated by \citet{Glass1998}. 
	
	In addition to the $LSP$ symmetries discussed above, the \rmat calculations were also performed for the higher total angular 
	momenta of the \electron-Li system up to $L=5$ in both singlet and triplet. Corresponding phase shifts (not shown in this 
	paper) are generally smaller than those for $L<3$ presented above and there are no resonances in the energy interval below 
	the \threes threshold. However, their inclusion in the calculation of the cross sections is necessary to achieve the 
	convergence and agreement with the experimental results.
	\subsection{Cross sections}
\begin{figure}
		\includegraphics{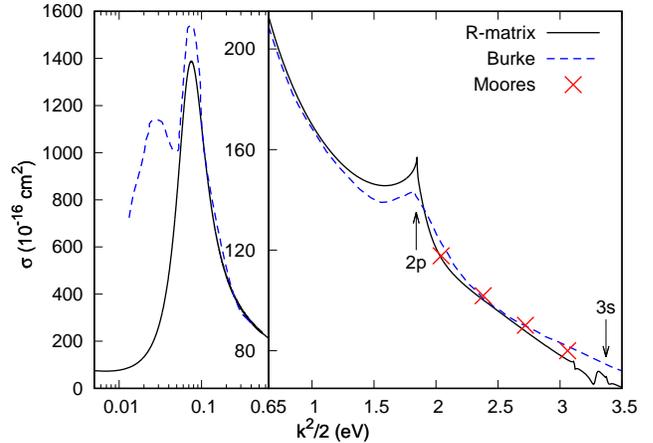}
		\caption{Cross section of the elastic electron collisions with lithium calculated using the \rmat method as a function of 
		the kinetic energy of the colliding electron (solid black line) and its comparison with the results of the CC 
		calculations \cite{Burke1969} (blue dashed line) as well as with the theoretical results published by \citet{Moores1986} 
		($\times$). The low-energy range is presented using the logarithmic energy scale.}
		\label{fig:cselex}
	\end{figure}

	All the scattering cross sections presented in this section are averaged over the initial spin states and summed over the 
	final spin states \cite{Nesbet1980book}. The integral cross sections for the elastic scattering calculated using the \rmat 
	method are plotted in \figab\ref{fig:cselex}.
	The dominant structure in the cross sections is the narrow peak located at 75~meV that corresponds to the low-lying \pothree 
	resonance discussed above (see \figab\ref{fig:phs3po} and \tabab\ref{tab:phsparams}). The change of the \sethree phase shift 
	from positive to negative values shown in the inset of \figab\ref{fig:phs3se} is reflected in the elastic cross section as 
	the Ramsauer-Townsend minimum located around 7~meV. As can be seen in \figab\ref{fig:cselex}, the resonance peak calculated 
	in 
	this work is in good agreement with that published by \citet{Burke1969}. However, their calculation yields one more 
	peak at lower energy. It is most likely an artifact caused by the non-orthogonality of the continuum wave functions on the 
	target orbitals \cite{Norcross1969} that is particularly problematic in the \sethree scattering.
	
	Another sharp peak in the elastic cross section occurs at the energy 1.84~eV where the \twop-channel opens. It is a
	consequence of the \deone resonance (see the corresponding phase shifts in \figab\ref{fig:phs13de}) and very pronounced 
	threshold behavior in the \poone and \pothree symmetries (Wigner cusp). In order to obtain converged cross sections in this 
	energy region, it is necessary to include all the total angular momenta of the \electron-Li system up to $L=4$ for both 
	singlet and triplet configurations. The difference of the cross sections obtained from the \rmat calculations discussed in 
	this work and 
	those reported by \citet{Burke1969} corresponds to the discrepancy of the phase shifts discussed above. 
	\figab\ref{fig:cselex} also shows the comparison with the elastic cross section calculated by \citet{Moores1986} who  
	utilized the CC approach involving five lowest states of the target. These results are in slightly better 
	agreement with the \rmat cross sections than those by \citet{Burke1969} obtained using more limited CC expansion, 
	although the elastic cross section calculated by \citet{Moores1986} decreases with the energy more slowly than our results 
	presented here.
	
	Although the \seone and \pothree resonances located close to the \threes-threshold are narrow (see 
	\tabab\ref{tab:phsparams}), 
	our results plotted in \figab\ref{fig:cselex} show that they do not dramatically change the magnitude of the cross sections. 
	Due to a sparse energy grid at which \citet{Moores1986} evaluated the cross sections, these resonances do not appear in that 
	theoretical study.
	
\begin{figure}
		\includegraphics{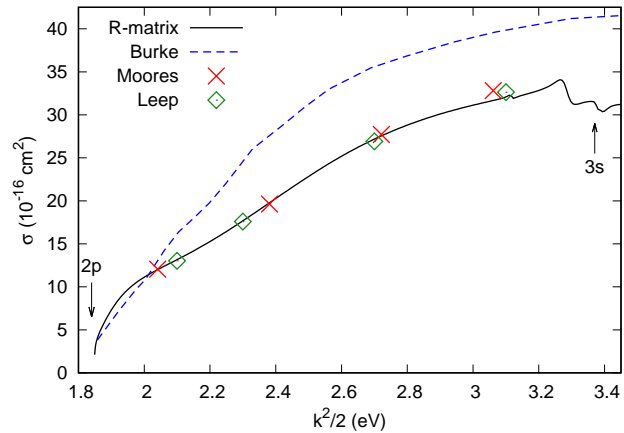}
		\caption{Cross sections of the $2s\to2p$ electronic excitation of lithium by electron impact at energies between the \twop 
		and \threes threshold. The solid black line represents the \rmat calculation presented in this work, the blue dashed line 
		is 
		the 
		result obtained by \citet{Burke1969}. Theoretical cross section by \citet{Moores1986} ($\times$) and experimental data by 
		\citet{Leep1974} ($\diamond$) are plotted for comparison.}
		\label{fig:csexcit}
	\end{figure}

	As can be seen in \figab\ref{fig:csexcit}, our \rmat calculations including the higher angular momenta and target orbitals 
	with higher energies yield lower cross sections for the electronic excitation to the \twop state than the CC 
	expansion of \citet{Burke1969}.
	The excellent agreement with the results of \citet{Moores1986} and with the experimental data of \citet{Leep1974} shows that 
	the five-states CC expansion \cite{Moores1986} provides converged and quantitatively accurate results of the electronic 
	excitation 
	in the energy range between the \twop and \threes threshold.
	
\begin{figure}
		\includegraphics{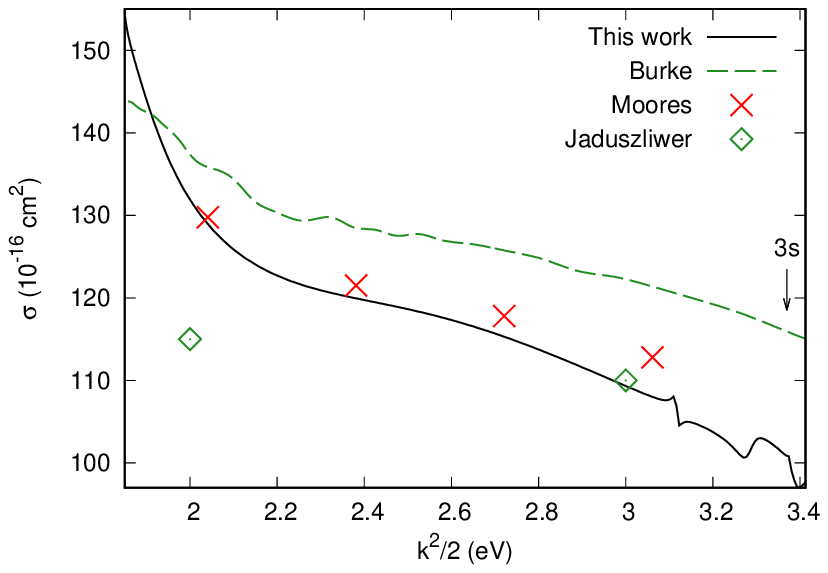}
		\caption{Total cross section for the \electron-Li collisions above the \twop threshold as sum of the elastic and 
		$2s\to2p$ excitation cross sections plotted in \figab\ref{fig:cselex} and \figab\ref{fig:csexcit}, respectively. The 
		solid black line represents the results of the 
		\rmat calculations, the blue dashed line is the cross section calculated by \citet{Burke1969}. The results by 
		\citet{Moores1986} are represented by the red crosses ($\times$) and the green diamonds ($\diamond$) are the 
		experimental values measured by \citet{Jaduszliwer1981}.}
		\label{fig:cstot}
	\end{figure}

	The sum of the elastic and $2s\to2p$ excitation cross section between the \twop and \threes thresholds is plotted in 
	\figab\ref{fig:cstot}.
	Present results are lower than the total cross section reported by \citet{Burke1969} and slightly lower than that published 
	by \citet{Moores1986}. Out of all the three theoretical results, present \rmat cross sections are closest to the experimental 
	data 
	observed by \citet{Jaduszliwer1981} in the energy region below the \threes threshold. \figab\ref{fig:cstot} shows that the 
	experimental value at energy of 3~eV is in excellent agreement with the \rmat 
	calculations presented here. It is not straightforward to address the difference that can be seen at the energy of 2~eV. It 
	is well 
	known that the crossed-beam scattering experiments become more challenging at lower collision energies. The energy profile of 
	the incident electron beam becomes broader and it is more difficult to form a well focused beam.
	
	\section{Conclusions}
	The aim of the present study is to provide accurate \abini data for the \electron-Li scattering that
	can be used for modeling ultra-long-range Rydberg molecules containing the Li atom. In the first step,
	the atomic core of Li$^+$ is replaced by a model potential whose parameters are fitted to accurately reproduce
	the atomic excitation energies. By very extensive size of the one-electron basis set
	we attempted to eliminate any impact of this basis on the model potential, i.e. the functional
	form of the potential should be considered independent of the basis set.
	
	The optimized model potential is then used in the following two-electron \rmat method developed
	by the authors \cite{Tarana2016} to compute the phase shifts for the total angular momentum up to
	$L=5$ and for the collision energies up to the \threes excitation threshold.
	The phase shifts up to $L=2$, in both singlet and triplet channels,
	are discussed in detail, while the higher $L$-values are
	only used to obtain converged integral cross sections presented in this work.
	
	Phase shifts for $L=0$ (\seone and \sethree symmetries) agree well with previous calculations
	of \citet{Norcross1971} and \citet{Burke1969} up to the first excitation threshold.
	The low-energy tail of the latter data required a correction to properly incorporate 
	orthogonality of the continuum states for the \sethree symmetry and it was
	carried out by \citet{Norcross1971}. The obtained low-energy MERT parameters can be considered
	as refinements of those published by \citet{Norcross1971}. For higher collision energies, our
	results deviate from those of \citet{Burke1969}. Moreover, we report a narrow 3$s^2$ Feshbach
	resonance at 3.117 eV located in \seone symmetry.
	
	Phase shifts for $L=1$ are dominated by the \poone resonance that was very well represented
	in calculations of \citet{Burke1969} and omitted by \citet{Norcross1971}. The narrow character
	of the resonance allowed us to fit the phase shift as a sum of the threshold-law phase and
	the Breit-Wigner formula. Resulting MERT parameter $B_3$ disagrees with the one provided
	by \citet{Norcross1971} as the latter was determined from higher collision energies for
	whose the \poone resonance at 62 meV does not exist.
	
	The resonance that can be seen in \deone phase shifts is wider when compared to those
	in lower total angular momenta $L$. Strong background together with the 
	presence of the 2$p$ excitation threshold provides very difficult situation for determination
	of the resonance parameters. Therefore, they are not given in the present study. Moreover,
	our results start to deviate from calculations of \citet{Burke1969} already well
	below the 2$p$ excitation threshold.
	
	The elastic as well as $2s\to2p$ inelastic integral cross sections were calculated for the energies below the \threes 
	excitation threshold.
	At very low energies, our results show
	the Ramsauer-Townsend minimum at 7~meV whereas in the calculations of \citet{Burke1969},
	this minimum is disturbed by a presence of an additional low-energy peak. Above the 2$p$ excitation
	threshold, the total cross section of \citet{Burke1969} becomes gradually higher than in the present calculations, mainly due 
	to 
	higher $2s \to 2p$ electronic excitation
	cross sections. Our results in this case agree well with those by \citet{Moores1986}.
	Furthermore, out of all three calculations, the present data exhibit the best agreement with the experimental total cross 
	sections \cite{Jaduszliwer1981}.
	
	The lack of any experimental data on the \electron-Li scattering for collision energies
	under 2~eV strongly underlines a necessity for the accurate theoretical results that
	could be utilized in design of ultra-cold molecular experiments \cite{Schmid2018}.
	Moreover, we believe that the techniques reported here for the construction of the
	model potential replacing the atomic core can be also used for the heavier alkali
	atoms.
	
	\begin{acknowledgments}
		M.T. acknowledges the support of the Czech Science Foundation (Project No. P203/17-26751Y). The contributions of R.\v{C}. 
		were supported by the Czech Science Foundation (Grant No. 18-02098S).
	\end{acknowledgments}
%

\end{document}